\documentclass[final]{svjour2}%
\usepackage{graphicx}
\usepackage{rotating}
\usepackage{amssymb}
\usepackage{mathptmx}
\usepackage[numbers]{natbib}
\usepackage{amsmath}
\usepackage{amsfonts}%
\providecommand{\U}[1]{\protect\rule{.1in}{.1in}}
\makeatletter
\journalname{Journal of Low Temperature Physics}
\bibpunct{}{}{,}{s}{}{,}
\begin{document}

\title{Pair excitations and parameters of state of imbalanced Fermi gases at finite temperatures}
\author{S. N. Klimin$^{1}$, J. Tempere$^{1,2}$, and Jeroen P. A. Devreese$^{1}$}

\institute{1:Theorie van Kwantumsystemen en Complexe Systemen, Universiteit
Antwerpen\\ Universiteitsplein 1, B-2020 Antwerpen, Belgium\\
\email{jacques.tempere@ua.ac.be}
\\2: Lyman Laboratory of Physics, Harvard University, Cambridge\\
Massachusetts 02138, USA}

\date{\today}
\maketitle

\begin{abstract}
The spectra of low-lying pair excitations for an imbalanced two-component
superfluid Fermi gas are analytically derived within the path-integral
formalism taking into account Gaussian fluctuations about the saddle point.
The spectra are obtained for nonzero temperatures, both with and without
imbalance, and for arbitrary interaction strength. On the basis of the pair
excitation spectrum, we have calculated the thermodynamic parameters of state
of cold fermions and the first and second sound velocities. The parameters of
pair excitations show a remarkable agreement with the Monte Carlo data and
with experiment.

PACS numbers: 03.75.Ss, 05.30.Fk, 03.75.Lm

\end{abstract}

\section{Introduction \label{sec:intro}}

Recent experimental breakthroughs in the manipulation of ultracold Bose and
Fermi gases have opened new prospects for advancing many-body physics
\cite{Bloch2008}. The dimensionality of these gases can be controlled with
optical lattices, and the interaction strength can be tuned using Feshbach
resonances. The experimental control over geometry and interactions in
ultracold atomic gases has turned these systems into powerful quantum
simulators that can test and generalize many-body theories originally
developed for solid state systems. Recently, much attention has been paid to
ultracold atomic gases with strong interactions because of their possible
relation to some striking natural phenomena including high-temperature
superconductors and neutron stars \cite{Petick}.

In particular, great experimental and theoretical effort has been devoted to
the study of superfluidity arising from pairing in ultracold Fermi gases
\cite{Radzihovsky}. Specifically, the effect of population imbalance (between
the pairing partners) on the superfluid pairing mechanism is a topic of
current investigation. Of no less interest in the context of high-temperature
superconductivity is the study of the crossover between the
Bardeen-Cooper-Schrieffer (BCS) and the molecular Bose-Einstein condensation
(BEC) regimes.

In order to describe the superfluid phase transition as well as the
broken-symmetry phase below a critical temperature $T_{c}$, several methods
based on the $T$-matrix approach have been developed
\cite{NSR,deMelo1993,Pieri,Chen,Combescot,Haussmann,Diener2008}. Amongst
those, the Nozi\`{e}res--Schmitt-Rink (NSR) theory \cite{NSR} and its
path-integral reformulation \cite{deMelo1993} have been very successful and
remain widely used. Here, we use an improved version of the NSR scheme, that
we will denote as the Gaussian pair fluctuation theory (GPF) \cite{Drum,Drum2}%
, and that effectively works both at low temperatures and above $T_{c}$.

As shown in Ref. \cite{Salasnich2010}, the thermodynamic properties of the
superfluid Fermi gas at sufficiently low temperatures can be derived within a
simple model using the spectrum of low-lying elementary excitations. In that
model, the thermodynamics of the superfluid Fermi gas \cite{Salasnich2010} is
based on the fermion-boson model with phenomenological parameters whose values
are determined from the Monte Carlo calculations \cite{MC1,MC2,MC3,Bulgac} (in
the limit of zero temperature). The model exploited in Ref.
\cite{Salasnich2010} describes well the thermodynamic properties of a balanced
unitary Fermi gas at low temperatures. Moreover it was demonstrated
\cite{Taylor2007} that in the BEC limit, the imbalanced Fermi superfluid
indeed reduces to a simple Bose-Fermi mixture of Bose-condensed molecules and
unpaired fermions. The goal of the present work is to extend these results to
non-zero temperatures, and to arbitrary scattering lengths.

For this purpose, we calculate the parameters needed for the fermion-boson
model from the NSR and/or GPF theories. Our results are compared with the
Monte Carlo data at unitarity. We analytically derive the spectra of low-lying
elementary excitations for an imbalanced Fermi gas in 3D at finite
temperatures in the whole range of the BCS-BEC\ crossover. These spectra are
obtained using the path-integral representation \cite{deMelo1993} of the NSR
theory extended to imbalanced Fermi gases \cite{PRA2008,PRB2008} as well as
the GPF approach \cite{Drum,Drum2}. Using the obtained spectra of the
elementary excitations, thermodynamic parameters such as the internal energy,
the chemical potential, the first and second sound velocities, are calculated.

\section{Formalism \label{sec:theory}}

\subsection{Path-integral GPF approach for imbalanced Fermi gases}

We consider a two-component Fermi gas within the path-integral approach. The
path-integral formulation \cite{deMelo1993} of the NSR scheme has been
extended in Refs. \cite{PRA2008,PRB2008} to the case of unequal `spin up' and
`spin down' populations of fermions. In the present work, the treatment of the
imbalanced Fermi gas is performed using the NSR scheme \cite{PRA2008,PRB2008}
and its improved version, the GPF theory \cite{Drum} extended to the
imbalanced case.

The thermodynamic parameters of the imbalanced Fermi gas are completely
determined by the thermodynamic potential $\Omega$ of the grand-canonical
ensemble. The thermodynamic potential $\Omega$, the same as in Refs.
\cite{PRA2008,PRB2008}, is the sum of the saddle-point thermodynamic potential
$\Omega_{sp}$ and the fluctuation contribution $\Omega_{fl}$. These
thermodynamic potentials are provided, respectively, by the zeroth-order and
quadratic terms of the expansion of the Hubbard-Stratonovich pair-field action
around the saddle point.

The saddle-point thermodynamic potential for the imbalanced Fermi gas with
$s$-wave pairing is \cite{PRA2008}
\begin{equation}
\Omega_{sp}=-V\int\frac{d\mathbf{k}}{\left(  2\pi\right)  ^{3}}\left[
\frac{1}{\beta}\ln\left(  2\cosh\beta\zeta+2\cosh\beta E_{\mathbf{k}}\right)
-\xi_{\mathbf{k}}-\frac{\Delta^{2}}{2k^{2}}\right]  -V\frac{\Delta^{2}}{8\pi
a_{s}} \label{Wsp}%
\end{equation}
where $V$ is the system volume, $\beta$ is the inverse to the thermal energy
$k_{B}T$, $\Delta$ is the amplitude of the gap parameter, $a_{s}$ is the
scattering length, $\xi_{\mathbf{k}}=k^{2}-\mu$ is the fermion energy, and
$E_{\mathbf{k}}=\sqrt{\xi_{\mathbf{k}}^{2}+\Delta^{2}}$ is the Bogoliubov
excitation energy. The chemical potentials of imbalanced fermions are
expressed through the averaged chemical potential $\mu=\left(  \mu_{\uparrow
}+\mu_{\downarrow}\right)  /2$ and the chemical potential imbalance
$\zeta=\left(  \mu_{\uparrow}-\mu_{\downarrow}\right)  /2$. We choose the
units with $\hbar=1$, the fermion mass $m=1/2$, and the Fermi energy
$E_{F}\equiv\hbar^{2}\left(  3\pi^{2}n\right)  ^{2/3}/\left(  2m\right)  =1$
($n$ is the total fermion density). The fluctuation contribution to the
thermodynamic potential $\Omega_{fl}$ is the same as in Refs.
\cite{PRA2008,PRB2008}.

The gap parameter is found from the gap equation minimizing the saddle-point
thermodynamic potential,%
\begin{equation}
\left.  \dfrac{\partial\Omega_{sp}(T,\mu,\zeta;\Delta)}{\partial\Delta
}\right\vert _{T,\mu,\zeta}=0
\end{equation}
from which we can extract $\Delta(T,\mu,\zeta)$. For an imbalanced gas, the
saddle-point thermodynamic potential can have two minima: one at $\Delta=0$
and one at $\Delta\neq0$. This can result in a first-order superfluid phase
transition \cite{Bedaque}. With our notation, we emphasize that the
thermodynamic potential is a function of $T,\mu,\zeta$ (and actually $V$, but
this dependency drops out). However, $\Delta$ is treated as an additional
parameter on which the thermodynamic potential depends. It is this treatment
of $\Delta$ as an additional parameter (in the broken-symmetry phase with
$\Delta\neq0$) that leads to a distinction between the NSR approach and the
GPF approach. When calculating the gap equation one should use%
\begin{align}
n &  =-\left.  \dfrac{\partial\Omega\left(  T,\mu,\zeta;\Delta\right)
}{\partial\mu}\right\vert _{T,\zeta,\Delta}-\left.  \dfrac{\partial\Omega
_{fl}\left(  T,\mu,\zeta;\Delta\right)  }{\partial\Delta}\right\vert
_{T,\zeta,\mu}\left.  \dfrac{\partial\Delta(T,\mu,\zeta)}{\partial\mu
}\right\vert _{T,\zeta},\\
\delta n &  =-\left.  \dfrac{\partial\Omega\left(  T,\mu,\zeta;\Delta\right)
}{\partial\zeta}\right\vert _{T,\mu,\Delta}-\left.  \dfrac{\partial\Omega
_{fl}\left(  T,\mu,\zeta;\Delta\right)  }{\partial\Delta}\right\vert
_{T,\zeta,\mu}\left.  \dfrac{\partial\Delta(T,\mu,\zeta)}{\partial\zeta
}\right\vert _{T,\mu}.
\end{align}
In the standard NSR approach the last terms (involving the derivatives of
$\Delta$) are omitted. The GPF approach suggested in Refs. \cite{Drum,Drum2}
takes into account the additional derivatives for the balanced case, and
corrects the NSR densities for changes in $\Delta$ as $\mu$ and $\zeta$ are
varied. The GPF method presented here is the path-integral formulation of the
GPF theory of Refs. \cite{Drum,Drum2} extended to the imbalanced case. As
shown in Ref. \cite{Drum2}, the GPF theory provides the best overall agreement
of its analytic results with experiment and with Monte Carlo data, except in
close vicinity to $T_{c}$.

The GPF corrections are not present when one uses the saddle-point
approximation and calculates%
\begin{align}
n_{sp} &  =-\left.  \dfrac{\partial\Omega_{sp}\left(  T,\mu,\zeta
;\Delta\right)  }{\partial\mu}\right\vert _{T,\zeta,\Delta},\\
\delta n_{sp} &  =-\left.  \dfrac{\partial\Omega_{sp}\left(  T,\mu
,\zeta;\Delta\right)  }{\partial\zeta}\right\vert _{T,\mu,\Delta}.
\end{align}
However, the correction terms will be relevant for the calculation of the
fluctuation contributions, $n_{fl}=n-n_{sp}$ and $\delta n_{fl}=\delta
n-\delta n_{sp}$. These fluctuation contributions to the density $n_{fl}$ and
$\delta n_{fl}$ are given by the same expressions as in Ref. \cite{PRA2008}
\begin{align}
n_{fl} &  =-\int\frac{d\mathbf{q}}{\left(  2\pi\right)  ^{3}}\left[  \frac
{1}{\pi}\int_{-\infty}^{\infty}\operatorname{Im}\frac{J\left(  \mathbf{q}%
,\omega+i\gamma\right)  }{e^{\beta\left(  \omega+i\gamma\right)  }-1}%
d\omega+\frac{1}{\beta}\sum_{n=-n_{0}}^{n_{0}}J\left(  \mathbf{q},i\nu
_{n}\right)  \right]  ,\label{nfl}\\
\delta n_{fl} &  =-\int\frac{d\mathbf{q}}{\left(  2\pi\right)  ^{3}}\left[
\frac{1}{\pi}\int_{-\infty}^{\infty}\operatorname{Im}\frac{K\left(
\mathbf{q},\omega+i\gamma\right)  }{e^{\beta\left(  \omega+i\gamma\right)
}-1}d\omega+\frac{1}{\beta}\sum_{n=-n_{0}}^{n_{0}}K\left(  \mathbf{q},i\nu
_{n}\right)  \right]  .\label{dnfl}%
\end{align}
Here, $n_{0}$ is an arbitrary positive integer, and the parameter $\gamma$
lies between two bosonic Matsubara frequencies $\nu_{n_{0}}<\gamma<\nu
_{n_{0}+1}$, $\nu_{n}\equiv2\pi n/\beta$. The spectral functions $J\left(
\mathbf{q},z\right)  $ and $K\left(  \mathbf{q},z\right)  $ of complex
frequency $z$ are%
\begin{align}
J\left(  \mathbf{q},z\right)   &  =\frac{M_{1,1}\left(  \mathbf{q},-z\right)
\frac{\partial M_{1,1}\left(  \mathbf{q},z\right)  }{\partial\mu}%
-M_{1,2}\left(  \mathbf{q},-z\right)  \frac{\partial M_{1,2}\left(
\mathbf{q},z\right)  }{\partial\mu}}{M_{1,1}\left(  \mathbf{q},z\right)
M_{1,1}\left(  \mathbf{q},-z\right)  -M_{1,2}^{2}\left(  \mathbf{q},z\right)
},\label{FJ}\\
K\left(  \mathbf{q},z\right)   &  =\frac{M_{1,1}\left(  \mathbf{q},-z\right)
\frac{\partial M_{1,1}\left(  \mathbf{q},z\right)  }{\partial\zeta}%
-M_{1,2}\left(  \mathbf{q},-z\right)  \frac{\partial M_{1,2}\left(
\mathbf{q},z\right)  }{\partial\zeta}}{M_{1,1}\left(  \mathbf{q},z\right)
M_{1,1}\left(  \mathbf{q},-z\right)  -M_{1,2}^{2}\left(  \mathbf{q},z\right)
},\label{FK}%
\end{align}
where $M_{j,k}\left(  \mathbf{q},z\right)  $ are the matrix elements of the
pair field propagator. The matrix elements $M_{j,k}\left(  \mathbf{q}%
,z\right)  $ are given by the expressions \cite{PRA2008}
\begin{align}
M_{1,1}\left(  \mathbf{q},z\right)   &  =\int\frac{d\mathbf{k}}{\left(
2\pi\right)  ^{3}}\left\{  \frac{1}{2k^{2}}+\frac{X\left(  E_{\mathbf{k}%
}\right)  }{2E_{\mathbf{k}}}\left[  \frac{\left(  z-E_{\mathbf{k}}%
+\varepsilon_{\mathbf{k}+\mathbf{q}}\right)  \left(  E_{\mathbf{k}%
}+\varepsilon_{\mathbf{k}}\right)  }{\left(  z-E_{\mathbf{k}}+E_{\mathbf{k}%
+\mathbf{q}}\right)  \left(  z-E_{\mathbf{k}}-E_{\mathbf{k}+\mathbf{q}%
}\right)  }\right.  \right.  \nonumber\\
&  \left.  \left.  -\frac{\left(  z+E_{\mathbf{k}}+\varepsilon_{\mathbf{k}%
+\mathbf{q}}\right)  \left(  E_{\mathbf{k}}-\varepsilon_{\mathbf{k}}\right)
}{\left(  z+E_{\mathbf{k}}-E_{\mathbf{k}+\mathbf{q}}\right)  \left(
z+E_{\mathbf{k}+\mathbf{q}}+E_{\mathbf{k}}\right)  }\right]  \right\}
-\frac{1}{8\pi a_{s}},\label{M11}%
\end{align}%
\begin{align}
M_{1,2}\left(  \mathbf{q},z\right)   &  =-\Delta^{2}\int\frac{d\mathbf{k}%
}{\left(  2\pi\right)  ^{3}}\frac{X\left(  E_{\mathbf{k}}\right)
}{2E_{\mathbf{k}}}\left[  \frac{1}{\left(  z-E_{\mathbf{k}}+E_{\mathbf{k}%
+\mathbf{q}}\right)  \left(  z-E_{\mathbf{k}}-E_{\mathbf{k}+\mathbf{q}%
}\right)  }\right.  \nonumber\\
&  \left.  +\frac{1}{\left(  z+E_{\mathbf{k}}-E_{\mathbf{k}+\mathbf{q}%
}\right)  \left(  z+E_{\mathbf{k}}+E_{\mathbf{k}+\mathbf{q}}\right)  }\right]
.\label{M22}%
\end{align}
Here, the distribution function $X\left(  E_{\mathbf{k}}\right)  $ is%
\begin{equation}
X\left(  E_{\mathbf{k}}\right)  =\frac{\sinh\beta E_{\mathbf{k}}}{\cosh\beta
E_{\mathbf{k}}+\cosh\beta\zeta}.\label{X}%
\end{equation}
The derivatives in (\ref{FJ}) and (\ref{FK}) within the GPF scheme are
determined as mentioned above -- taking into account a variation of the gap
parameter. Within the NSR scheme, these derivatives are calculated assuming
$\Delta$ to be an independent variational parameter.

The equation of state of the imbalanced Fermi gas is thus determined as a
joint solution of the saddle-point gap equation and the number equations
accounting for Gaussian fluctuations. Within both the NSR and GPF schemes, the
Gaussian fluctuations do not feed back into the saddle-point gap equation.

\subsection{Low-lying pair excitations}

In order to obtain the spectrum of low-lying pair excitations for the
imbalanced Fermi gas, we perform the long-wavelength and low-energy expansion
of the matrix elements $M_{j,k}\left(  \mathbf{q},z\right)  $ as proposed in
Ref. \cite{Diener2008}. We take into account the terms up to quadratic order
in powers of $q$ and $z$, and find:%
\begin{align}
M_{1,1}\left(  \mathbf{q},z\right)   &  \approx A+Bq^{2}+Cz+Qz^{2},\nonumber\\
M_{1,2}\left(  \mathbf{q},z\right)   &  \approx D+Eq^{2}+Hz^{2}.
\label{series1}%
\end{align}

The coefficients of the expansion (\ref{series1}) are derived
straightforwardly. After some algebra, we arrive at their expression through
the integrals:%
\begin{align}
A  &  =\frac{1}{2\pi^{2}}\int k^{2}dk\left(  \frac{1}{2k^{2}}-\frac{E_{k}%
^{2}+\xi_{k}^{2}}{4E_{k}^{3}}X\left(  E_{\mathbf{k}}\right)  -\frac{\Delta
^{2}}{4}\frac{X^{\prime}\left(  E_{k}\right)  }{E_{k}^{2}}\right)  -\frac
{1}{8\pi a_{s}},\nonumber\\
B  &  =\frac{1}{48\pi^{2}}\int k^{2}dk\frac{2E_{k}^{4}k^{2}-3E_{k}^{4}\xi
_{k}+9\xi_{k}^{3}E_{k}^{2}+14E_{k}^{2}\xi_{k}^{2}k^{2}-20\xi_{k}^{4}k^{2}%
}{E_{k}^{7}}X\left(  E_{\mathbf{k}}\right) \nonumber\\
&  +\frac{\Delta^{2}}{24\pi^{2}}\int k^{2}dk\frac{1}{E_{k}^{4}}\left(
\frac{3\xi_{k}\left(  E_{k}^{2}-2\xi_{k}k^{2}\right)  }{E_{k}^{2}}X^{\prime
}\left(  E_{k}\right)  \right. \nonumber\\
&  \left.  +\frac{6\xi_{k}^{2}k^{2}-E_{k}^{2}\left(  3\xi_{k}+2k^{2}\right)
}{2E_{k}}X^{\prime\prime}\left(  E_{k}\right)  -\frac{2k^{2}\xi_{k}^{2}}%
{3}X^{(3)}\left(  E_{k}\right)  \right)  ,\nonumber\\
C  &  =-\frac{1}{8\pi^{2}}\int k^{2}dk\frac{\xi_{k}}{E_{k}^{3}}X\left(
E_{\mathbf{k}}\right)  -\frac{\Delta^{2}}{8\pi^{2}}\int k^{2}dk\frac
{X^{\prime}\left(  E_{k}\right)  }{\xi_{k}E_{k}^{2}},\nonumber\\
D  &  =\frac{\Delta^{2}}{8\pi^{2}}\int k^{2}dk\frac{X\left(  E_{\mathbf{k}%
}\right)  }{E_{\mathbf{k}}^{3}}-\frac{\Delta^{2}}{8\pi^{2}}\int k^{2}%
dk\frac{X^{\prime}\left(  E_{k}\right)  }{E_{k}^{2}},\nonumber
\end{align}%
\begin{align}
E  &  =\frac{\Delta^{2}}{48\pi^{2}}\int k^{2}dk\frac{20\xi_{k}^{2}k^{2}%
-E_{k}^{2}\left(  9\xi_{k}+6k^{2}\right)  }{E_{k}^{7}}X\left(  E_{\mathbf{k}%
}\right) \nonumber\\
&  +\frac{\Delta^{2}}{24\pi^{2}}\int k^{2}dk\frac{1}{E_{k}^{4}}\left(
\frac{E_{k}^{2}\left(  3\xi_{k}+2k^{2}\right)  -6\xi_{k}^{2}k^{2}}{2E_{k}^{2}%
}\left[  2X^{\prime}\left(  E_{k}\right)  -E_{k}X^{\prime\prime}\left(
E_{k}\right)  \right]  \right. \nonumber\\
&  \left.  -\frac{2k^{2}\xi_{k}^{2}}{3}X^{(3)}\left(  E_{k}\right)  \right)
,\nonumber\\
Q  &  =-\frac{1}{32\pi^{2}}\int k^{2}dk\frac{E_{k}^{2}+\xi_{k}^{2}}{E_{k}^{5}%
}X\left(  E_{\mathbf{k}}\right)  ,\nonumber\\
H  &  =\frac{\Delta^{2}}{32\pi^{2}}\int_{0}^{\infty}k^{2}dk\frac{X\left(
E_{\mathbf{k}}\right)  }{E_{k}^{5}}. \label{coefs}%
\end{align}

The low-lying pair excitations correspond to the poles of the spectral
functions (\ref{FJ}) and (\ref{FK}). Therefore the dispersion equation for the
energies of the pair excitations $\omega=\omega_{q}$ is
\begin{equation}
M_{1,1}\left(  \mathbf{q},\omega\right)  M_{1,1}\left(  \mathbf{q}%
,-\omega\right)  -M_{1,2}^{2}\left(  \mathbf{q},\omega\right)  =0.
\label{disp}%
\end{equation}
This equation is solved expanding $\omega_{q}^{2}$ up to the terms of the
order of $q^{4}$. We then obtain the energies $\omega_{q}$ in a form
structurally similar to the collective excitations in Ref.
\cite{Salasnich2010}:%
\begin{equation}
\omega_{q}=\sqrt{c^{2}q^{2}+\lambda q^{4}}, \label{spectrum}%
\end{equation}
where the parameters $c$ and $\lambda$ are related to the coefficient of the
expansion (\ref{series1}) as follows,%
\begin{align}
c  &  =\left(  \frac{2A\left(  B-E\right)  }{C^{2}+2A\left(  H-Q\right)
}\right)  ^{1/2},\label{c1a}\\
\lambda &  =\frac{C^{2}\left(  B-E\right)  \left(  4A\left(  BH-EQ\right)
+C^{2}\left(  B+E\right)  \right)  }{\left(  C^{2}+2A\left(  H-Q\right)
\right)  ^{3}}. \label{lambda}%
\end{align}

For small pair momentum $q$, the energy of the pair excitation becomes linear
in the momentum. Thus $\omega_{q}$ at small $q$ represents a
Bogoliubov--Anderson mode, which is gapless in accordance with the
Nambu--Goldstone theorem. The parameter $c$ has the dimensionality of
velocity. In the zero-temperature limit for a balanced gas, all fermions are
in the superfluid state, and the velocity parameter for the pair excitations
tends to the first sound velocity for the whole fermion system. The so-called
gradient parameter $\lambda$ provides a growth of kinetic energy due to a
spatial variation of the density \cite{Salasnich2008,S2}.

The pair excitation spectra obtained in the present work generalize the
long-wavelength expansion of Ref. \cite{Diener2008} to the case of non-zero
temperatures and unequal `spin-up' and `spin-down' fermion populations.%

\begin{figure}
[ptb]
\begin{center}
\includegraphics[
height=4.1788in,
width=2.949in
]%
{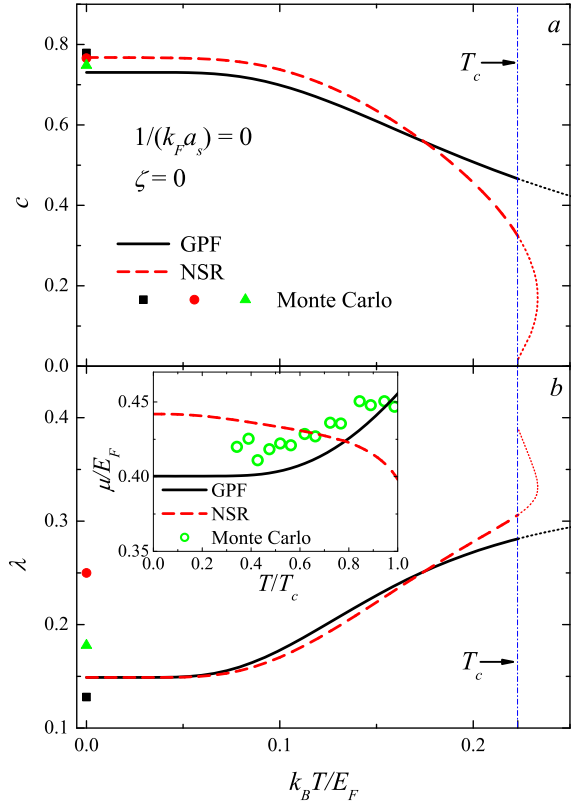}%
\caption{(color online) (\emph{a}) Sound velocity parameter of the
Bogoliubov--Anderson mode $c$ for cold fermions in 3D as a function of
temperature in the unitarity regime and for $\zeta=0$. The parameters of state
are determined taking into account the fluctuations in different ways: within
the path-integral GPF formalism (solid curve) and within the NSR scheme
(dashed curve). The symbols show the results of the different Monte Carlo
calculations \cite{MC1,MC1a,MC2,MC2a,MC3}. (\emph{b}) The parameter $\lambda$
as a function of temperature in the same regime. The dot-dashed vertical line
indicates the critical temperature of the superfluid phase transition. The
thin dotted curves show the formal solutions for the parameter $c$ and
$\lambda$ within the path-integral GPF and NSR approaches above $T_{c}$.
\emph{Inset}: Chemical potential calculated within the path-integral GPF and
NSR approaches compared with the Monte Carlo data from Ref. \cite{Bulgac}.}%
\end{center}
\end{figure}

In Fig. 1, the parameters $c$ and $\lambda$ characterizing pair excitations
are plotted as a function of temperature for the balanced Fermi gas in the
unitarity regime ($1/a_{s}=0$). The parameters of the equation of state for a
given temperature are determined from the joint solution of the gap and number
equations taking into account fluctuations in the number equation. The
fluctuation contributions to the fermion density are calculated within the
path-integral GPF and NSR schemes.

As found in Ref. \cite{Drum}, the NSR approach becomes inaccurate in the
vicinity of the critical temperature $T_{c}$. It was also shown that the NSR
scheme reveals a re-entrant behavior of the parameters in the state above
$T_{c}$, leading to an artificial first-order superfluid phase transition
\cite{Fukushima2007}. The re-entrant behavior of the parameters $c$ and
$\lambda$ obtained in the NSR approach is clearly seen in Fig. 1. The critical
temperature $T_{c}$ for the balanced gas$,$ indicated by a dash-dotted line in
Fig. 1, is the same within the NSR and GPF approaches. However, the GPF method
leads to better results with respect to NSR for the broken-symmetry phase.
This can be seen, for example, in the inset of Fig. 1 where we plot the
chemical potential as a function of temperature below $T_{c}$ calculated
within the NSR and path-integral GPF approaches and compared with the Monte
Carlo results of Ref. \cite{Bulgac}.

In the zero-temperature limit, the sound velocity parameter $c$ for the pair
excitations obtained within both the path-integral GPF and NSR approaches
exhibits an excellent agreement with the numerical results obtained using
different Monte Carlo algorithms \cite{MC1,MC1a,MC2,MC2a,MC3}. Also the
gradient parameter $\lambda$ at zero temperature lies within the range of the
values of $\lambda$ obtained in Refs. \cite{Salasnich2010,S2} as the best
fitting parameters for the ground state energy of fermions compared with Monte
Carlo data. This agreement demonstrates the accuracy of the present approach
for the broken-symmetry phase of cold fermions.%

\begin{figure}
[h]
\begin{center}
\includegraphics[
height=7.5674cm,
width=10.5328cm
]%
{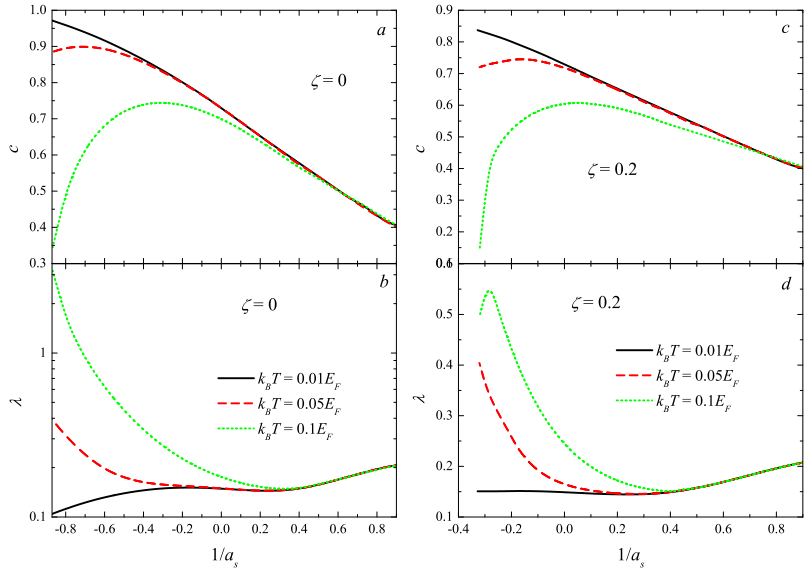}%
\caption{(color online) Sound velocity parameter (\emph{a},\emph{~c}) and
parameter $\lambda$ (\emph{b},\emph{~d}) for the Bogoliubov--Anderson mode as
a function of the inverse scattering length for $k_{B}T=0.01E_{F}$ (solid
curves), for $k_{B}T=0.05E_{F}$ (dashed curves) and for $k_{B}T=0.1E_{F}$
(dotted curves) for a balanced Fermi gas (\emph{a},\emph{~b}) and at the
chemical potential imbalance $\zeta=0.2$ (\emph{c},\emph{~d}) in the unitarity
regime.}%
\end{center}
\end{figure}

In Fig. 2, we plot the parameters $c$ and $\lambda$ as a function of the
inverse scattering length $1/a_{s}$ for the balanced gas (the left-hand
panels) and at the chemical potential imbalance $\zeta=0.2$ (the right-hand
panels). At $k_{B}T=0.01E_{F}$ the sound velocity parameter monotonously
decreases with increasing $1/a_{s}$. For non-zero temperatures, $c$ exhibits a
maximum, which shifts to higher coupling strengths for higher temperatures.
The parameter $\lambda$ at finite temperatures has a minimum, which almost
vanishes in the zero-temperature limit. In the weak-coupling regime, both $c$
and $\lambda$ are sensitive to temperature. When moving towards the
strong-coupling regime, $c$ and $\lambda$ gradually become almost independent
on $T$. The imbalance leads to the appearance of a critical inverse scattering
length such that for smaller $1/a_{s}$, there is no superfluid state (see, e.
g., Ref. \cite{PRB2008}).

\section{Parameters of state \label{sec:thermpars}}

\subsection{Thermodynamic functions}

Using the spectra of the elementary fermionic and pair excitations derived in
Sec. \ref{sec:theory}, we can obtain the thermodynamic functions of the
superfluid Fermi gas at finite temperature. In Ref. \cite{Salasnich2010}, a
similar description of the thermodynamic properties was performed for a
unitary balanced Fermi gas using the zero-temperature spectra of elementary excitations.

In Ref. \cite{Salasnich2010}, a pair excitation spectrum of the form of
expression (\ref{spectrum}) is used, where the zero-temperature sound velocity
$c$ is taken from the Monte Carlo data \cite{MC1,MC1a,MC2,MC2a,MC3} and the
gradient parameter $\lambda$ is determined from a fit of the thermodynamic
properties to the Monte Carlo results. In the present calculation, the pair
excitation spectra are obtained using the analytic path-integral GPF approach
without any fit.

In this section we consider the thermodynamic functions of the cold Fermi gas
within the model of fermionic and pair excitations. The grand-canonical
thermodynamic potential is the sum of the saddle-point and pair excitation
contributions%
\begin{equation}
\Omega=\Omega_{sp}+\Omega_{p}. \label{W}%
\end{equation}
The saddle-point thermodynamic potential $\Omega_{sp}$ is given by Eq.
(\ref{Wsp}). The contribution of pair excitations is \cite{Salasnich2010}%
\begin{equation}
\Omega_{p}=\frac{V}{\beta}\int\frac{d\mathbf{q}}{\left(  2\pi\right)  ^{3}}%
\ln\left(  1-e^{-\beta\omega_{q}}\right)  . \label{Wp}%
\end{equation}

The entropy $S$ is found using its relation to the grand-canonical
thermodynamic potential,%
\begin{equation}
S=-\left.  \frac{\partial\Omega}{\partial T}\right\vert _{V,\mu,\zeta
}.\label{S}%
\end{equation}
Using the thermodynamic potential $\Omega$ with (\ref{Wsp}) and (\ref{Wp}) we
find that the entropy is expressed as%
\begin{align}
S &  =\frac{V\beta^{2}}{3\pi^{2}}\int_{0}^{\infty}\frac{\xi_{k}}{E_{k}}%
\frac{E_{k}\left(  \cosh\beta E_{k}\cosh\beta\zeta+1\right)  -\zeta\sinh\beta
E_{k}\sinh\beta\zeta}{\left(  \cosh\beta\zeta+\cosh\beta E_{k}\right)  ^{2}%
}k^{4}dk\nonumber\\
&  +\frac{V}{2\pi^{2}}\int_{0}^{\infty}\left(  \beta\omega_{q}\frac
{e^{-\beta\omega_{q}}}{1-e^{-\beta\omega_{q}}}-\ln\left(  1-e^{-\beta
\omega_{q}}\right)  \right)  q^{2}dq.\label{S1}%
\end{align}
Finally, the internal energy $E$ of cold fermions is calculated using the
relation between $E$ and the grand-canonical thermodynamic potential,
$E=\Omega+TS+\mu N,$where $N$ is the total number of fermions.

In Fig. 3, the internal energy per particle for a unitary Fermi gas calculated
in different approaches is plotted as a function of temperature for the
broken-symmetry phase at $T\leqslant T_{c}$. The critical temperature
determined within the path-integral GPF model is the same as within NSR,
$T_{c}\approx0.225E_{F}/k_{B}$. The results of the present calculation within
the model of fermionic and low-lying pair excitations with parameters
determined using the path-integral GPF and NSR methods are shown with
short-dashed and dot-dashed curves, respectively. The other results
represented in Fig. 3 are (after Ref. \cite{Salasnich2010}): the internal
energy calculated within the low-temperature fermion-boson (FB) model
\cite{Salasnich2010}, and the result of the analytic model proposed by Bulgac,
Drut, and Magierski (BDM) \cite{Bulgac}. The analytic results are compared
with those of the Monte Carlo calculations from Ref. \cite{Bulgac} and with
the experimental data of Ref. \cite{Horikoshi} for a gas of $^{6}$Li atoms at unitarity.%

\begin{figure}
[h]
\begin{center}
\includegraphics[
height=6.882cm,
width=8.8853cm
]%
{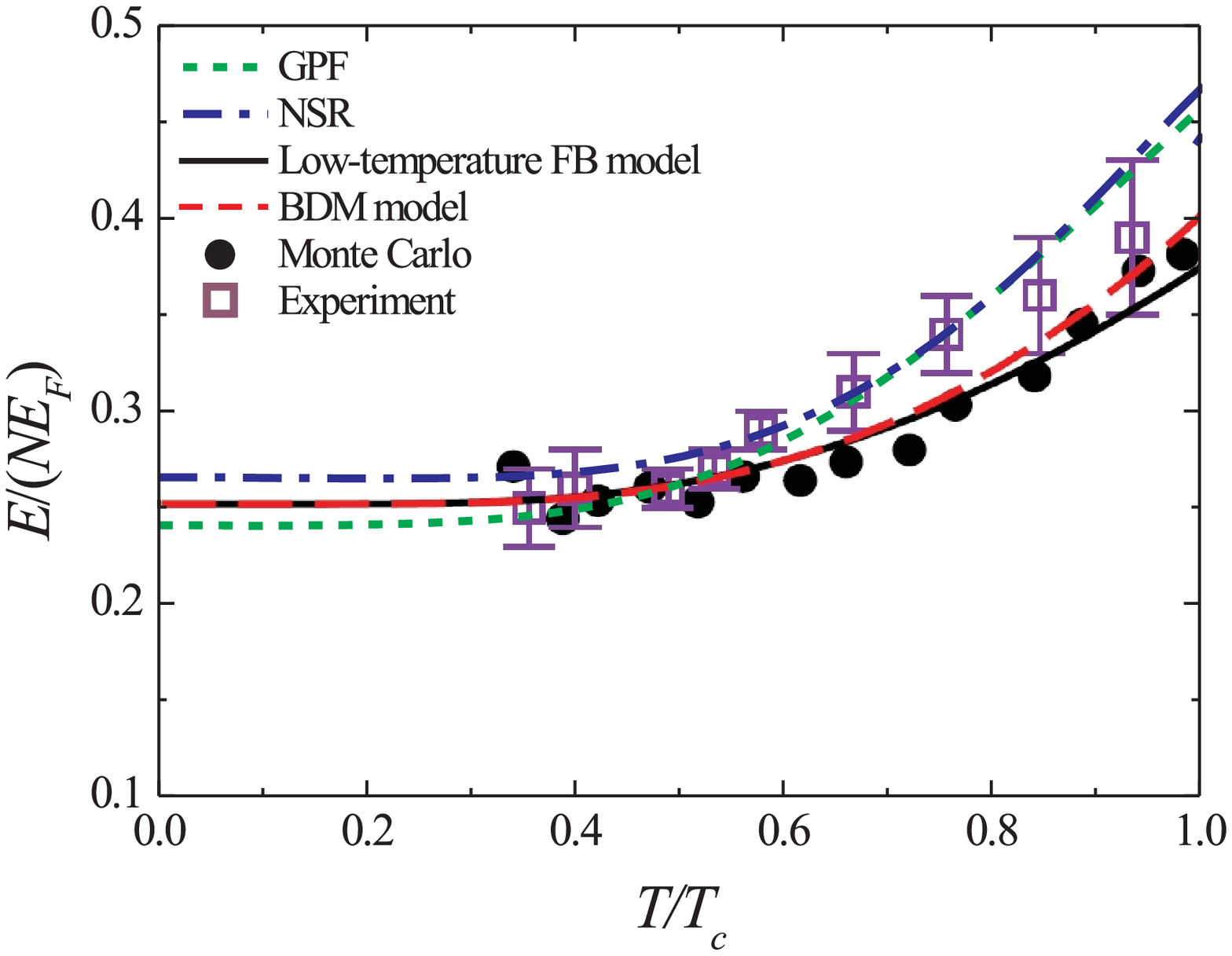}%
\caption{(color online) Internal energy per particle for a Fermi gas at
unitarity calculated within the model of fermionic and low-lying pair
excitations with parameters determined using the path-integral GPF method
(short-dashed curve) and the NSR theory (dot-dashed curve). The solid curve
shows the internal energy obtained within the low-temperature fermion-boson
model in Ref. \cite{Salasnich2010}. The dashed curve is the result of the BDM
model \cite{Bulgac}. The full dots represent the Monte Carlo simulations
\cite{Bulgac}. The squares are the experimental data \cite{Horikoshi}.}%
\end{center}
\end{figure}

As reported in Ref. \cite{Salasnich2010}, the low-temperature fermion-boson
model works well in the broken-symmetry phase where the internal energy
resulting from this model is close to the Monte Carlo data of Ref.
\cite{Bulgac}. The present study is performed with the parameters of the
elementary excitations obtained using the analytic approaches rather than a
fit to the Monte Carlo simulations. The internal energy calculated in the
present work with the parameters determined using the path-integral GPF method
is very close to the Monte Carlo results at low temperatures $T\lessapprox
0.55T_{c}$. Furthermore, our result is in good agreement with the experiment
\cite{Horikoshi} in the whole temperature range below $T_{c}$.

Neither the model of fermionic and pair excitations used in the present work,
nor the low-temperature fermion-boson model of Ref. \cite{Salasnich2010}
predicts the superfluid phase transition: the critical temperature $T_{c}$ is
determined within the path-integral GPF method before the long-wavelength
expansion is performed in Sec. \ref{sec:theory}. However, the latter model can
describe well the broken-symmetry phase of cold fermionic atoms.

\subsection{Superfluid density}

The total density of cold fermions within the model of fermionic and pair
excitations is given by the sum of fermion and boson contributions%
\begin{equation}
n=\frac{1}{2\pi^{2}}\int_{0}^{\infty}k^{2}dk\left(  1-\frac{\xi_{k}}{E_{k}%
}X_{k}\right)  +\frac{1}{\pi^{2}}\int_{0}^{\infty}q^{2}dq\frac{1}%
{e^{\beta\omega_{q}}-1}.\label{n}%
\end{equation}
The total density is a sum of the normal and superfluid densities
$n=n_{n}+n_{s}$. The superfluid density $n_{s}$, as well as the total density,
is constituted by the saddle-point result for an imbalanced Fermi gas and the
contribution due to the pair excitations,
\begin{align}
n_{s} &  =\frac{1}{2\pi^{2}}\int_{0}^{\infty}\left(  1-\frac{\xi_{k}}{E_{k}%
}X_{k}-k^{2}Y_{k}\right)  k^{2}dk\nonumber\\
&  +\frac{1}{\pi^{2}}\int_{0}^{\infty}\left(  \frac{1}{e^{\beta\omega_{q}}%
-1}-\frac{\beta}{3}q^{2}\frac{e^{-\beta\omega_{q}}}{\left(  e^{-\beta
\omega_{q}}-1\right)  ^{2}}\right)  q^{2}dq,\label{nsp}%
\end{align}
where the function $Y_{k}$ is given by%
\[
Y_{k}\equiv\frac{\partial X_{k}}{\partial E_{k}}=\beta\frac{\cosh\beta
E_{k}\cosh\beta\zeta+1}{\left(  \cosh\beta E_{k}+\cosh\beta\zeta\right)  ^{2}}%
\]
(see the analogous expression for the Fermi gas in 2D in Ref. \cite{TKD2009}).
In the balanced case, the superfluid density becomes equivalent to the
corresponding expression of Ref. \cite{Salasnich2010}, but with other values
of the parameters of the pair excitations, as discussed above and shown in Fig.1.%

\begin{figure}
[h]
\begin{center}
\includegraphics[
height=6.7524cm,
width=8.6701cm
]%
{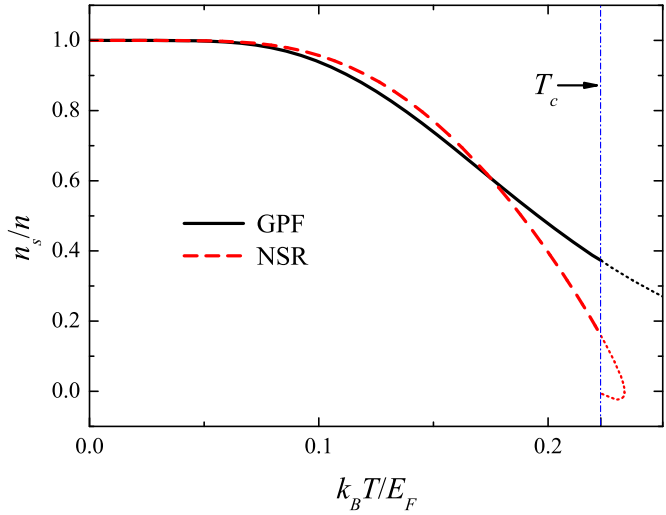}%
\caption{(color online) Superfluid density for a Fermi gas at unitarity
calculated within the model of fermionic and pair excitations with parameters
determined using the path-integral GPF method (solid curve) and the NSR theory
(dashed curve). Thin dotted curves correspond to the temperature range above
$T_{c}$.}%
\end{center}
\end{figure}

In Fig. 4, the superfluid density divided by the total density calculated
within the model of fermionic and pair excitations is plotted as a function of
temperature. As seen from the figure, the superfluid density calculated using
the parameters of the pair excitations obtained within the NSR model exhibits
a re-entrant behavior above $T_{c}$ similarly to the sound velocity parameter
in Fig. 1. The analogous bend-over of the superfluid density above $T_{c}$ was
reported in the full NSR approach in Ref. \cite{Fukushima2007}.

\subsection{Sound velocities}

We consider the sound propagation in a superfluid Fermi gas using the approach
of the two-fluid hydrodynamics \cite{Landau,Khalatnikov} in the same way as in
Ref. \cite{Salasnich2010}. The first sound velocity $u_{1}$ in the two-fluid
hydrodynamics is determined by the formula%
\begin{equation}
u_{1}=\left(  2\left.  \frac{\partial P}{\partial n}\right\vert _{\bar{S}%
,V}\right)  ^{1/2}, \label{u1}%
\end{equation}
with the entropy per particle $\bar{S}=S/N$. The pressure is proportional to
the grand-canonical thermodynamic potential: $P=-\Omega/V$. We adopt the
expression \cite{Taylor2009}
\begin{equation}
\left.  \frac{\partial P}{\partial n}\right\vert _{\bar{S},V}=\frac{5}{3}%
\frac{P}{n} \label{dP}%
\end{equation}
and use the grand-canonical thermodynamic potential given by Eq. (\ref{W})
with (\ref{Wsp}) and (\ref{Wp}).

The second sound velocity $u_{2}$ characterizes the temperature waves in which
the motion of the normal and superfluid fractions is out-of-phase. It is
determined by the formula%
\begin{equation}
u_{2}=\left(  \frac{2\bar{S}^{2}}{\left.  \frac{\partial\bar{S}}{\partial
T}\right\vert _{N,V}}\frac{n_{s}}{n_{n}}\right)  ^{1/2}. \label{u2}%
\end{equation}

The formulae (\ref{u1}) and (\ref{u2}) are valid as far as the first-sound and
second-sound modes are decoupled. Following Ref. \cite{Salasnich2010}, we
assume that the above condition is fulfilled for a cold Fermi gas.%

\begin{figure}
[h]
\begin{center}
\includegraphics[
height=6.7524cm,
width=8.6152cm
]%
{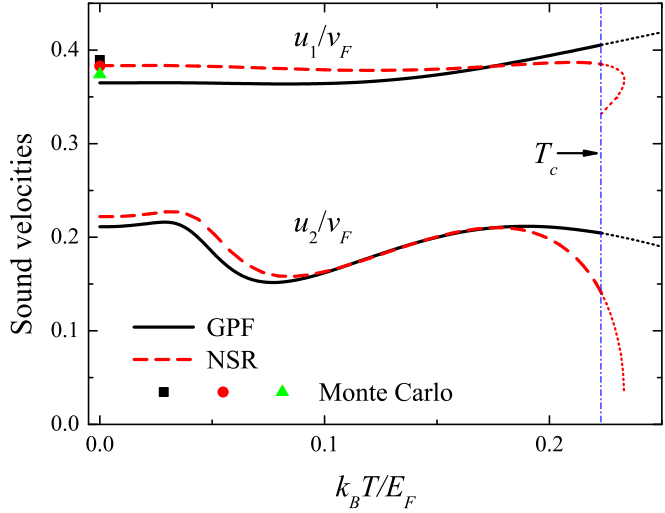}%
\caption{(color online) First and second sound velocities for a unitary Fermi
gas calculated within the model of fermionic and pair excitations with
parameters determined using the path-integral GPF method (solid curves) and
the NSR theory (dashed curves). The symbols show the Monte Carlo results for
the first sound velocity. Thin dotted curves correspond to the temperature
range above $T_{c}$.}%
\end{center}
\end{figure}

In Fig. 5, the first and second sound velocities (divided by the
zero-temperature Fermi velocity $v_{F}$) are plotted as a function of
temperature. They are calculated within the model of fermionic and pair
excitations using the parameters of pair excitations determined by the
path-integral GPF and NSR methods.

The first and second sound velocities obtained using the model of fermionic
and pair excitations are in a reasonable agreement with the results of the
analysis based on the full NSR thermodynamics \cite{Taylor2009,Arahato}. In
the zero-temperature limit, the first sound velocity tends to the same limit
as the sound velocity parameter $c$ for pair excitations, which is extremely
close to the Monte Carlo data \cite{MC1,MC1a,MC2,MC2a,MC3}.

\section{Conclusions \label{sec:comclusions}}

In the present work, the GPF modification \cite{Drum,Drum2} of the NSR scheme
has been formulated in the path-integral representation and extended to the
case of imbalanced fermions. Within this path-integral GPF approach, we have
analytically derived the spectra of low-lying pair excitations of the
imbalanced Fermi gas with $s$-wave pairing at finite temperatures and
extracted the parameters $c$ and $\lambda$ for the pair excitations from these
results. Using these spectra, the finite-temperature thermodynamics of the
Fermi gas in the superfluid state has been analyzed. The obtained internal
energy demonstrates a good agreement with the Monte Carlo results and is
remarkably close to the experimental data for the Fermi gas at unitarity. The
zero-temperature value of the first sound velocity is in a good agreement with
the results of the Monte Carlo simulations. The present method allows us to
obtain the spectra of the elementary excitations and, consequently, the
thermodynamic parameters of the state for an arbitrary scattering length, at
non-zero temperatures, and for non-zero imbalance.

\begin{acknowledgement}
The authors gratefully acknowledge discussions with L. Salasnich. This work was
funded by the Fonds voor Wetenschappelijk Onderzoek-Vlaanderen (FWO-V)
projects G.0356.06, G.0370.09N, G.0180.09N, and G.0365.08. JPAD acknowledges
financial support in the form of a Ph.D. Fonds voor Wetenschappelijk Onderzoek-Vlaanderen (FWO-V).
\end{acknowledgement}

\end{document}